\documentclass[dvips]{article}

\usepackage{icrc2011}

\title{Monitoring of bright, nearby Active Galactic Nuclei with the MAGIC telescopes}
\newcommand{\etal}{\MakeLowercase{\textit{et al.}}} 
\shorttitle{Wagner, R. M., Backes, M., Satalecka, K. \etal, AGN Monitoring with MAGIC}

\authors{Robert WAGNER$^{1,2}$, Michael BACKES$^{3}$, Konstancja SATALECKA$^{4}$, Giacomo BONNOLI$^{5}$, Marlene DOERT$^{3}$, Burkhard STEINKE$^{1}$, Nikola STRAH$^{3}$, Tomislav TERZIC$^{6}$, Diego TESCARO$^{7}$, and Malwina UELLENBECK$^{3}$ for the MAGIC Collaboration}
\afiliations{
$^{1}$Max-Planck-Institut f\"ur Physik, F\"ohringer Ring 6, D-80805 M\"unchen, Germany\\
$^{2}$Excellence Cluster ``Universe'', TU M\"unchen, Boltzmannstra\ss e, D-85748 Garching b. M\"unchen, Germany\\
$^{3}$Technische Universit\"at Dortmund, D-44221 Dortmund, Germany\\
$^{4}$DESY, Platanenallee 6, D-15738 Zeuthen, Germany\\
$^{5}$INAF National Institute for Astrophysics, I-00136 Rome, Italy\\
$^{6}$Croatian MAGIC Consortium, Rudjer Boskovic Institute, Universities of Rijeka and Split, HR-10000 Zagreb, Croatia\\
$^{7}$IFAE, Edifici Cn., Campus UAB, E-08193 Bellaterra, Spain
}
\email{robert.wagner@mpp.mpg.de, michael.backes@physik.tu-dortmund.de, konstancja.satalecka@desy.de}

\abstract{Observations and detections of Active Galactic Nuclei (AGN) by
Cherenkov telescopes are often triggered by information about high flux states
in other wavelength bands. To overcome this bias, the VHE gamma-ray telescope
MAGIC has conducted dedicated monitoring observations of nearby AGN since 2006.
Three well established, TeV-bright blazars were selected to be observed
regularly: Mrk 421, Mrk 501, and 1ES1959+650.
The goals of these observations are to obtain an unbiased distribution of flux
states shedding light on the duty cycle of AGN, to investigate potential
spectral changes during periods of different source activity, and to correlate
the results with multiwavelength observations. Also clues on a potential
periodic behavior of the sources might be drawn from a study of the obtained
lightcurves. By testing predictions of theoretical models, like, e.g., the
correlation between the TeV flux level and the peak frequency predicted in SSC
models, monitoring deepens our knowledge about the acceleration and emission
processes in AGN.
The status and results of the MAGIC AGN monitoring program will be presented.}

\begin{document}
\maketitle

\section{Introduction}
\label{sec:intro}

By far, most of the known extragalactic emitters of very high energy (VHE, E$>$100\,GeV) gamma-rays are Active Galactic Nuclei (AGN). Among those, the largest subclass is comprised by blazars, being characterized by the relativistic jet pointing towards the observer. They show non-thermal continuum emission ranging from radio to VHE gamma-rays, covering an energy range of more than~15 orders of magnitude. The spectral energy distribution (SED) of this emission typically shows a two hump structure with one hump ranging from radio to X-rays and the second one peaking in the GeV to TeV range. The emission is typically highly variable in all wavebands and on timescales ranging from minutes~\cite{MAGIC:501flare, HESS:2155flare} to years~\cite{periods:OJ287,periods:optical_plates}.
Although theo\-retical models, generally, can explain quite well the shape of the observed blazar SEDs, the nature of the underlying acceleration mechanism is still under debate. The question whether electrons or hadrons are causing the electromagnetic emission in blazars is far from being setteled.
In leptonic models the low energy hump of the SEDs is caused by synchrotron radiation of a population of highly relativistic electrons. The same electron population may afterwards interact either with those synchrotron photons (Synchrotron Self Compton, SSC)~\cite{model:ssc} or an external photon field (External Compton, EC)~\cite{model:ec} via inverse Compton scattering, accounting for the high energy emission. These models describe most of the observed data very well and can reasonably explain even the shortest variability timescales.

Hadronic emission models are generally more complicated as they also feature leptonic emission processes for secondary leptons, which also in this scenario are dominantly contributing to the low energy hump via synchrotron emission. The high energy bump, in turn, is caused by either proton synchrotron emission (Proton Synchrotron Blazar, Synchrotron Mirror Model)~\cite{model:psb,model:smm}, or by the decay of neutral pions, stemming from interactions of protons with internal or external photons fields or among themselves~\cite{model:pic}. These scenarios are not only able to explain the SED shapes reasonably well, but are also capable of naturally explaining ``orphan flares''. These are enhancements of the high energy flux which are not accompanied by a simultaneous enhancement in the low energy emission and have been observed for the blazars 1ES\,1959+650~\cite{orphan:1959}, Mkn\,421~\cite{orphan:421}, and recently also for Mkn\,501~\cite{orphan:501}. These observations cannot be explained by leptonic scenarios featuring only one emission zone, where a correlated behavior of the low and high energy emission is expected.
Another feature of hadronic emission models is the prediction of high energy neutrino emission~\cite{nu:mannheim, nu:halzen, nu:marlene}.

Imaging Atmospheric Cherenkov Telescopes (IACTs) of the latest generation like Cangaroo\footnote{\texttt{http://icrhp9.icrr.u-tokyo.ac.jp/}}, H.E.S.S.\footnote{\texttt{http://www.mpi-hd.mpg.de/hfm/HESS/}}, MAGIC\footnote{\texttt{http://wwwmagic.mpp.mpg.de/}}, and VERITAS\footnote{\texttt{http://veritas.sao.arizona.edu/}} are probing deeply the VHE region. Being much more sensitive than satellite or water Cherenkov experiments, like, e.g., {\it Fermi}-LAT\footnote{\texttt{http://fermi.gsfc.nasa.gov/}} and MILAGRO\footnote{\texttt{http://umdgrb.umd.edu/cosmic/milagro/}}, respectively, IACTs suffer from their extremely limited fields of view, compared to the former. Thus, instead of all-sky surveys IACTs are  used for deep single source exposures, leading to a dependency on external triggers for the observation of already known sources.

\section{AGN Monitoring}
\label{sec:monitoring}
To overcome the limitations of observations triggered by high flux states in other wavelengths, monitoring observations which are independent of the source state are mandatory. Only by such observations, it is possible to obtain an unbiased distribution of flux states and, by this, deduce the duty cycle and thus the flaring state probability of the observed sources. This is especially important for studying the sources' behavior in a multi-wavelength or even multi-messenger context: on the one hand, this holds true because correlations between fluxes in different flux states may directly shed light on the underlying acceleration mechanism, as stated in Section~\ref{sec:intro}. On the other hand, only a robust estimation of the flaring probabilities may allow for an estimation of the statistical significance of a possible correlation of blazar flares with possible extra terrestrial neutrinos observed by IceCube~\cite{nu:konst}.
Furthermore, such observations allow to investigate the possibility of changes in the emission spectra depending on the flux level of the sources and to compare such findings with the predictions of the theoretical models. Another important aspect is the possibility to trigger Target of Opportunity~(ToO) observations of either telescopes in other wavelength bands, or even other IACTs. Especially for that reason, the VERITAS collaboration still operates the Whipple~10\,m telescope~\cite{monitoring:whipple}, and even a worldwide network of Cherenkov telescopes conducting blazar monitoring is proposed~\cite{monitoring:dwarf}. Moreover, conducting an unbiased monitoring is obviously the only handle to the detection of ``orphan flares''.

\section{The MAGIC telescopes}
The MAGIC Telescopes are the largest IACTs for VHE gamma-ray astronomy, featuring two times~234\,m$^2$ mirror area. They are situated at~2.200\,m~a.s.l. in the Observatorio del Roque de los Muchachos of the European Northern Observatory on the Canary Island of La Palma. The first MAGIC telescope has been in scientific operation since~2004 and underwent a major upgrade in the beginning of~2007 when a~2\,GSamples/s FADC data acquisition system was installed~\cite{magic:mux}. With this high temporal resolution the influence of background photons could significantly be diminished and the separation of signal events from the hadronic background background events could be improved~\cite{magic:timing}. Both effects resulted in a sensitivity such that a source emitting a gamma-ray flux of~1.6\% of that of the Crab~Nebula with the same spectral behavior could be established with 5\,$\sigma$ significance within~50\,h of observation time. Thus, above~300\,GeV, a flux of~30\% of the Crab~Nebula flux could be detected within~30\,min~\cite{monitoring:09}. 
Furthermore, MAGIC can be operated under moderate moonlight and twilight conditions and despite the much higher amount of background photons these data can be processed with the standard software pipeline~\cite{magic:mars} without any further modifications~\cite{magic:moon}.
In~2009 the second telescope started scientific operation, leading to a sensitivity improvement of a factor of two in the whole energy range from 50\,GeV to several~TeV, improving at the same time also the energy resolution~\cite{magic:2}.


\section{Monitoring strategy}
According to the conditions stated in Section~\ref{sec:monitoring} observations have been scheduled in an unbiased way, not making use of~e.g. information from other wavelengths. They have been evenly distributed over the according observation periods. The sources chosen for the presented monitoring campaign are Mkn\,421, Mkn\,501, and 1ES\,1959+650: the former being on average the two VHE~brightest blazars, the latter one being especially interesting because of an huge increase in flux in~2002 when simultaneously even two neutrino events from the same direction have been observed by AMANDA, whereas this was not statistically significant~\cite{nu:bernadini}. The observations for the brighter sources had a typical length of 15-30\,min, lasting at least 30\,min for 1ES\,1959+650. As MAGIC is capable to observe during moderate moon and twilight conditions, a sizable amount of the monitoring has been conducted in such conditions.

\section{Results}
In the following we present preliminary results of the monitoring of nearby AGN with the MAGIC telescope during the complete phase of mono-scopic observations, i.e. from summer 2004 until summer 2009. Data taken under poor observation conditions have been rejected based e.g. on trigger rate, atmospheric transparency, and mean sky brightness measured as DC current of the Photomultipliers. Most of the observations have been performed in wobble-mode, i.e. with simultaneous determination of the flux of background events. All analysis steps have been verified on contemporary Crab Nebula data.

\begin{figure*}[th]
  \centering
  \includegraphics[width=6in,height=2.64in]{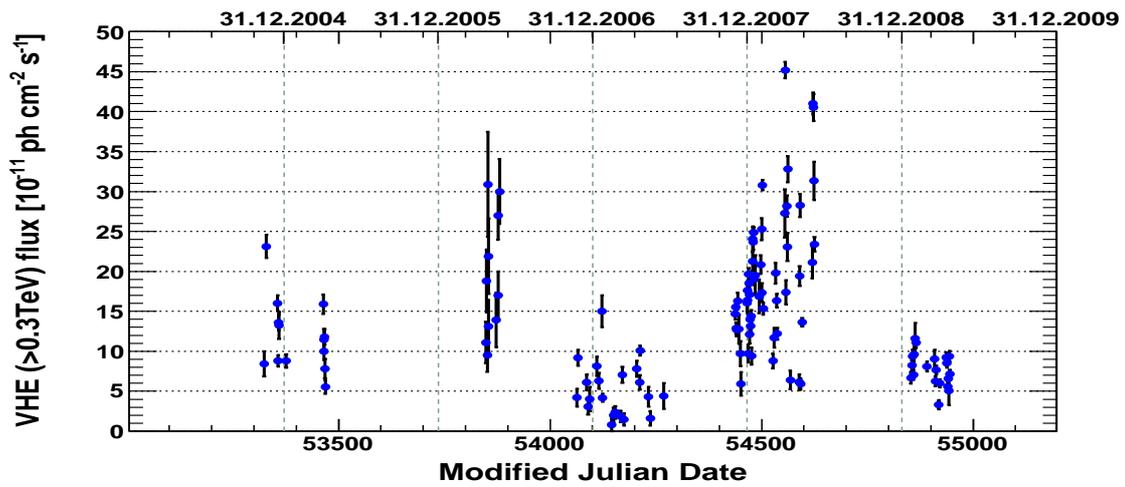}
  \caption{Preliminary MAGIC lightcurve of Mkn\,421 observed above 0.3\,TeV from~2004 until~2009. Observations less significant than 1\,$\sigma$ have been omitted.}
  \label{fig:421}
 \end{figure*}

\paragraph{Mkn\,421} is on average the brightest AGN in the VHE regime. It has been observed by MAGIC since~2004~\cite{mkn421:04}, showing remarkable flaring activity in~2006~\cite{mkn421:06flare} during a multiwavelength campaign~\cite{mkn421:06mwl} and in~2008~\cite{mkn421:08}, revealing a significant correlation of the VHE and X-ray emission~\cite{monitoring:09}. Altogether there have been~118 pointings of MAGIC resulting in the lightcurve depicted in Fig.~\ref{fig:421}.

\begin{figure*}[th]
  \centering
 \includegraphics[width=6in,height=2.64in]{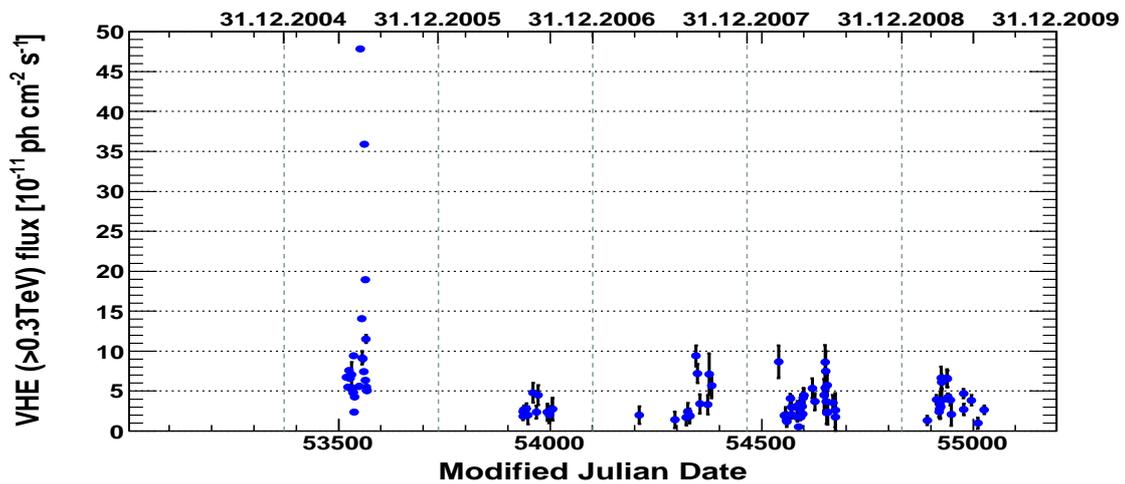}
  \caption{Preliminary MAGIC lightcurve of Mkn\,501 observed above 0.3\,TeV from~2005 until~2009. Observations less significant than 1\,$\sigma$ have been omitted.}
  \label{fig:501}
 \end{figure*}

\paragraph{Mkn\,501} was observed in an extreme outburst in 2005, showing a strong correlation of the spectral shape on the observed flux~\cite{MAGIC:501flare}. Since that time it has been observed with the MAGIC telescope, resulting in over~90\,hrs of observations on~103 days, presented in Fig.~\ref{fig:501}. Subsets of these data have already been published elsewhere~\cite{MAGIC:501flare, mkn501:06, mkn501:09m+v, mkn501:09fermi}.

\begin{figure*}[th]
  \centering
  \includegraphics[width=6in,height=2.64in]{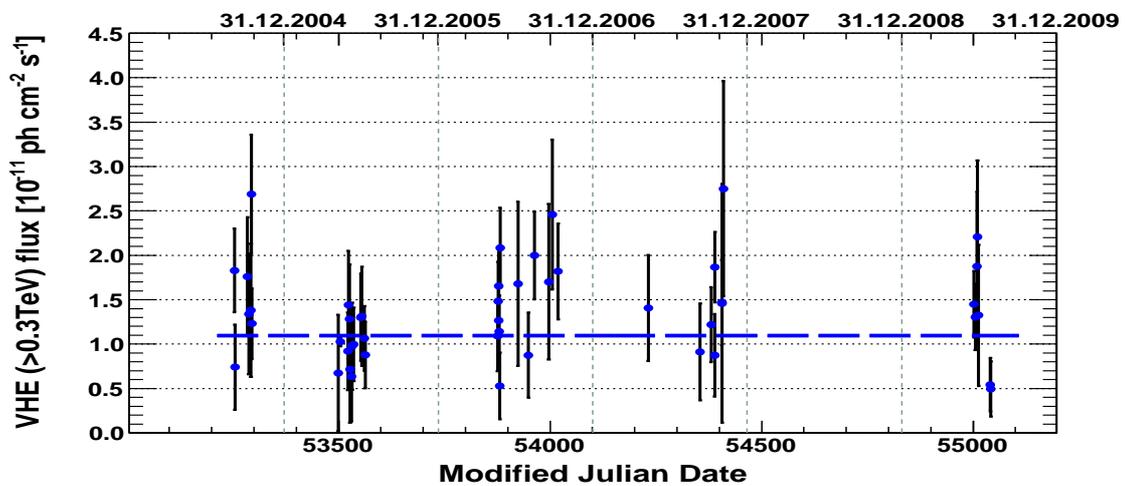}
  \caption{Preliminary MAGIC lightcurve of 1ES\,1959+650 observed above 0.3\,TeV from~2004 until~2009. Observations less significant than 1\,$\sigma$ have been omitted.}
  \label{fig:1959}
 \end{figure*}

\paragraph{1ES\,1959+650}
was one of the first objects observed by MAGIC in~2004, resulting in one of the very first MAGIC publications~\cite{1959:2004}. Since then a regular monitoring has been conducted~\cite{monitoring:07, monitoring:09, monitoring:09scinghe}, resulting in~$\sim$58\,hrs of useful data, spread over~47\,days with more than~1\,$\sigma$ detections as shown in Fig.~\ref{fig:1959}. Considering also the previous publications~\cite{1959:2004, 1959:2006}, no significant variations neither in flux nor in spectral shape have been observed.

\section{Conclusion \& Outlook}
We have presented preliminary results of the long-term monitoring campaign conducted with the first MAGIC telescope over six years. For subsets, significant correlations of the VHE and X-ray fluxes have been observed~\cite{monitoring:09scinghe} as well as dependence of the spectral slope on the flux state~\cite{MAGIC:501flare}. Further studies on the flux state distribution, the spectral slope dependence on the flux and correlations with other wavelengths are ongoing and will be presented elsewhere.

\paragraph{Acknowledgments}
We would like to thank the Instituto de Astrof\'{\i}sica de Canarias for the excellent working conditions at the Observatorio del Roque de los Muchachos in La Palma. The support of the German BMBF and MPG, the Italian INFN, the Swiss National Fund SNF, and the Spanish MICINN is gratefully acknowledged. This work was also supported by the Marie Curie program, by the CPAN CSD2007-00042 and MultiDark CSD2009-00064 projects of the Spanish Consolider-Ingenio 2010 programme, by grant DO02-353 of the Bulgarian NSF, by grant 127740 of the Academy of Finland, by the YIP of the Helmholtz Gemeinschaft, by the DFG Cluster of Excellence ``Origin and Structure of the Universe'', by the DFG Collaborative Research Centers SFB823/C4 and SFB876/C3, and by the Polish MNiSzW grant 745/N-HESS-MAGIC/2010/0.

\clearpage

\end{document}